\documentclass[12pt, prd, showpacs]{revtex4}
\usepackage{amssymb}
\usepackage{amsmath}

\input{tcilatex}

\begin{document}

\title{Near-horizon circular orbits and extremal limit for dirty rotating
black holes }
\author{O. B. Zaslavskii}
\affiliation{Department of Physics and Technology, Kharkov V.N. Karazin National
University, 4 Svoboda Square, Kharkov, 61077, Ukraine}
\affiliation{Institute of Mathematics and Mechanics, Kazan Federal University, 18
Kremlyovskaya St., Kazan 420008, Russia}
\email{zaslav@ukr.net }

\begin{abstract}
We consider generic rotating axially symmetric "dirty" (surrounded by
matter) black holes. Near-horizon circular equatorial orbits are examined in
two different cases of near-extremal (small surface gravity $\kappa $) and
exactly extremal black holes. This has a number of qualitative distinctions.
In the first case, it is shown that such orbits can lie as close to the
horizon as one wishes on suitably chosen slices of space-time when $\kappa
\rightarrow 0$. This generalizes observation of T.\ Jacobson Class. Quantum
Grav. 28 187001 (2011) made for the Kerr metric. If a black hole is extremal
($\kappa =0$), circular on-horizon orbits are impossible for massive
particles but, in general, are possible in its vicinity. The corresponding
black hole parameters determine also the rate with which a fine-tuned
particle on the noncircular near-horizon orbit asymptotically approaches the
horizon. Properties of orbits under discussion are also related to the Ba%
\~{n}ados-Silk-West effect of high energy collisions near black holes.
Impossibility of the on-horizon orbits in question is manifestation of
kinematic censorship that forbids infinite energies in collisions.
\end{abstract}

\maketitle

\section{Introduction}

Circular orbits near rotating black holes is an important issue in
astrophysics and are of interest theoretically on their own since they
possess a number of nontrivial properties. There are three main types of
such orbits -\ the photon orbit, the marginally bound and the marginally
stable ones. The latter is also called the innermost stable circular orbit
(ISCO). It was shown in a "classic" paper \cite{72} for the Kerr metric that
in the extremal limit, all three orbits share the same value of the
Boyer-Lindquist radial coordinate $r_{0}$ that coincides with the horizon
radius $r_{H}$. Hereafter, subscript "H" refers to the quantity taken on the
horizon. However, it was stressed in \cite{72} that it does not mean that
they coincide with the horizon. The radial coordinate becomes degenerate
there and is \ unsuitable for analysis. A more thorough inspection showed
that the proper distance between each orbit and the horizon and between
different orbits is not zero in the limit under discussion (moreover, the
distance between the marginally bound orbit and ISCO even diverges).

Meanwhile, some subtleties concerning the properties of these orbits
remained unnoticed until recently. As was pointed out in \cite{ted11}, the
very meaning of location of orbits depends crucially on what slice of
space-time is chosen for inspection. There are such slices that the proper
distance to the horizon tends to zero, so in the extremal limit these orbits
asymptotically approach the horizon.

Formally, $\lim_{\kappa \rightarrow 0}r_{0}=r_{H}$, where $r_{0}$ is the
radius of the orbit, $\kappa $ is the surface gravity ($\kappa =0$ for
extremal black holes). However, the situation when a black hole is extremal
from the very beginning should be considered separately. Time to time
debates on this issue have been continuing \cite{ind2}, \cite{m} in spite of
the fact that, clearly, a trajectory of a massive particle cannot lie on the
light-like surface. Detailed analysis of this circumstance was done in \cite%
{circkerr} for the Kerr metric. With the help of a coordinate system
suggested below, we carry out analysis for a generic axially symmetric
rotating black hole and show that the corresponding circular orbit is indeed
fake.

Although such orbits with $r_{0}=r_{H}$ are fake, we show that the values of
black hole parameters for which the radii of these orbits coincide with $%
r_{H}$ do have some physical meaning. This meaning is twofold. Circular
orbits with radii slightly above the horizon do exist, then in the main
approximation the black hole parameters under discussion are not arbitrary
but take some fixed values. On the other hand, these values also
characterize dynamics, when a particle moves along the trajectory that is
not exactly circular. Then, the rate with which a particle asymptocially
approaches the horizon changes when black hole angular momentum (or another
parameter) crosses these values in the space of parameters.

The aim of the present work is to give a general description of the
equatorial near-horizon orbits in question for a generic stationary
axially-symmetric black hole. We do not specify its metric and do not assume
it to be necessarily the Kerr or Kerr--Newman one. This deviation can be
attributed to matter that inevitably surrounds a black hole in astrophysics.
In this sense, a black hole is "dirty". In the equatorial plane, we
introduce a coordinate system that generalizes the version of the coordinate
system suggested\ for the Kerr metric in \cite{dor}. Hopefully, this system
can be of interest not only for the issue of near-horizon circular orbits
but can have some general value. We consider separately the extremal limits
of nonextremal black holes and extremal black holes that leads to
qualitatively different results.

The properties of near-horizon circular orbits are intimately related to the
so-called Ba\~{n}ados-Silk-West (BSW) effect \cite{ban}. According to this
effect, if two particles collide near a black \ hole, the energy in their
centre of mass $E_{c.m.}$ can be, under certain conditions, formally
unbound. The fact that the on-horizon orbit of a massive particle is fake
turns out to be intimately connected with impossibility to gain infinite $%
E_{c.m.}.$ This quantity remains finite in any act of collision, although it
can become as large as one likes (this can be called "the kinematic
censorship"). The relation between the high energy collisions and properties
of ISCO close to the horizon was considered first in \cite{circkerr} for the
near-extremal Kerr metric and generalized in \cite{circ} for dirty black
holes. In the present paper, we discuss the aforementioned relation both for
near-extremal and extremal black holes. We also give geometric explanation
to the high energy collisions of particles near ISCO. This can be considered
as a counterpart of geometric explanation given for the standard BSW effect 
\cite{cqg}.

We would like to stress that studies of motion of particles in the
background of the Kerr-Newman black hole has attracted interest during many
years until recently \cite{nar} - \cite{ruf3}. In general, details of motion
are quite complicated and depend crucially on the concrete properties of the
metric. In our approach, we restrict ourselves by the near-horizon region
only. This enables us to trace some features that rely on the properties of
the horizon in a model-independent way. This can be thought of as one more
manifestation of \ universality of black hole physics.

Throughout the paper, we put fundamental constants $G=c=1$.

\section{Coordinate system and equations of motion}

We consider the metric of the form

\begin{equation}
ds^{2}=-N^{2}dt^{2}+g_{\phi }(d\phi -\omega dt)^{2}+\frac{dr^{2}}{A}%
+g_{\theta }d\theta ^{2}\text{,}  \label{met}
\end{equation}%
where the metric coefficients do not depend on $t$ and $\phi $.
Correspondingly, the energy $E=-mu_{0}$ and angular momentum $L=mu_{\phi 
\text{ }}$are conserved. Here, $m\,\ $is the particle's mass$,u^{\mu }$ is
the four-velocity. In what follows, we restrict ourselves by motion in the
equatorial plane $\theta =\frac{\pi }{2}$ only. Then, it is convenient to
redefine the radial coordinate $r\rightarrow \rho $ in such a way that 
\begin{equation}
ds^{2}=-N^{2}dt^{2}+g_{\phi }(d\phi -\omega dt)^{2}+\frac{d\rho ^{2}}{N^{2}}.
\end{equation}

Equations of motion for a test particles moving along the geodesic read%
\begin{equation}
m\frac{dt}{d\tau }=\frac{X}{N^{2}}\text{,}  \label{t}
\end{equation}%
\begin{equation}
X\equiv E-L\omega \text{,}  \label{X}
\end{equation}%
\begin{equation}
m\frac{d\phi }{d\tau }=\frac{L}{g_{\phi }}+\frac{\omega X}{N^{2}}\text{,}
\label{phi}
\end{equation}%
where $\tau $ is the proper time.

The forward in time condition $\frac{dt}{d\tau }>0$ requires 
\begin{equation}
X\geq 0\text{,}  \label{ft}
\end{equation}%
where $X=0$ is possible on the horizon only, where $N=0$.

Using also the normalization condition $u_{\mu }u^{\mu }=-1$, one can infer
that%
\begin{equation}
m^{2}(\frac{d\rho }{d\tau })^{2}=Z^{2}\equiv X^{2}-N^{2}(\frac{L^{2}}{g}%
+m^{2})\text{.}  \label{ro}
\end{equation}

We also may introduce new (barred) coordinates according to%
\begin{equation}
dt=d\bar{t}-\frac{\chi d\rho }{N^{2}}\text{,}  \label{tt}
\end{equation}%
\begin{equation}
d\phi =d\bar{\phi}-\frac{wd\rho }{N^{2}},\rho =\bar{\rho}\text{,}
\end{equation}%
where the functions $\chi $ and $w$ depend on $\rho $ only. Then,%
\begin{equation}
d\phi -\omega dt=d\bar{\phi}-\omega d\bar{t}+\frac{\omega \chi -w}{N^{2}}%
d\rho ,
\end{equation}%
\begin{equation}
g_{\rho \rho }=\frac{1-\chi ^{2}}{N^{2}}+\frac{g_{\phi }(\omega \chi -w)^{2}%
}{N^{4}}\text{.}
\end{equation}%
We choose the functions $\chi $, $w$ to kill divergences in the metric
coefficient $g_{\rho \rho }$. To this end, we require%
\begin{equation}
\chi ^{2}=1-\mu N^{2}\text{,}  \label{hi}
\end{equation}%
\begin{equation}
\omega \chi -w=N^{2}h\text{,}  \label{w}
\end{equation}%
where $h(\rho )$ and $\mu (\rho )$ are bounded functions, $h(\rho _{H})\neq
0 $, $\mu (\rho _{H})\neq 0$. We obtain

\begin{equation}
d\phi -\omega dt=d\bar{\phi}-\omega d\bar{t}+hd\rho \text{.}
\end{equation}%
It is convenient to choose $\mu =1$, $h=-w$. Then,%
\begin{equation}
\chi =\sqrt{1-N^{2}\text{ }}\text{, }w\chi =\omega \text{.}  \label{ww}
\end{equation}

As a result,%
\begin{equation}
ds^{2}=-N^{2}d\bar{t}^{2}+g_{\phi }(d\bar{\phi}-\omega d\bar{t}%
)^{2}+(1+g_{\phi }w^{2})d\rho ^{2}-2g_{\phi }(d\bar{\phi}-\omega d\bar{t}%
)wd\rho +2\chi d\bar{t}d\rho .  \label{md}
\end{equation}

It can be also rewritten in the form%
\begin{equation}
ds^{2}=-d\bar{t}^{2}+(1+g_{\phi }w^{2})(d\rho +\chi d\bar{t}-\frac{g_{\phi }w%
}{1+g_{\phi }w^{2}}d\bar{\phi})^{2}+\frac{g_{\phi }}{1+g_{\phi }w^{2}}d\bar{%
\phi}^{2}\text{.}  \label{md2}
\end{equation}

For the energy and angular momentum we have%
\begin{equation}
\mathcal{E}\equiv \frac{E}{m}=\frac{\bar{E}}{m}=-g_{\bar{t}\bar{t}}\frac{d%
\bar{t}}{d\tau }+\omega g_{\phi }\frac{d\bar{\phi}}{d\tau }-(\omega g_{\phi
}w+\chi )\frac{d\rho }{d\tau }\text{,}  \label{E}
\end{equation}%
\begin{equation}
\mathcal{L}\equiv \frac{L}{m}=\frac{\bar{L}}{m}=g_{\phi }\frac{d\bar{\phi}}{%
d\tau }-\omega g_{\phi }\frac{d\bar{t}}{d\tau }-g_{\phi }w\frac{d\rho }{%
d\tau }.  \label{L}
\end{equation}%
These quantities can be also written as $\mathcal{E}=-u_{\mu }\xi ^{\mu }$, $%
\mathcal{L}=u_{\mu }\eta ^{\mu }$, where $\xi ^{\mu }$ and $\eta ^{\mu }$
are Killing vectors responsible for time translation and azimuthal rotation,
respectively, $u^{\mu }=\frac{dx^{\mu }}{d\lambda }$, $\lambda $ is the
affine parameter along the geodesic. Therefore, the expressions (\ref{E}), (%
\ref{L}) for $\mathcal{E}$ and $\mathcal{L}$ are valid both in the massive
and massless cases.

Let us consider motion in the inward direction, so $\frac{d\rho }{d\tau }<0$%
. It follows from (\ref{ro}) and (\ref{tt}) that%
\begin{equation}
m\frac{d\bar{t}}{d\tau }=\frac{X-\chi Z}{N^{2}}\text{,}  \label{tz}
\end{equation}%
\begin{equation}
\frac{d\bar{\phi}}{d\tau }=\frac{L}{mg_{\phi }}+\frac{\omega X-wZ}{mN^{2}}%
\text{.}  \label{phi1}
\end{equation}%
One finds from (\ref{tz})%
\begin{equation}
\bar{t}=-\int \frac{d\rho (X-\chi Z)}{ZN^{2}}\text{.}  \label{int}
\end{equation}%
If $X_{H}\neq 0$, near the horizon $Z\approx X_{H}+O(N^{2})$, so the
integral converges. Thus, in contrast to the time $t$ (analogue of the
Boyer-Lindquist time), the time $\bar{t}$ required to reach the horizon is
finite. Only for a special trajectory with $X_{H}=0$ (so-called critical
particles - see below) $\bar{t}$ is infinite.

In the particular case $L=0$, $E=m$ we have from (\ref{tz}) and (\ref{ro}), (%
\ref{ww}) or from (\ref{E}), (\ref{L}) that%
\begin{equation}
\frac{d\bar{t}}{d\tau }=1\text{, }\frac{d\bar{\phi}}{d\tau }=0\text{,}
\end{equation}%
so $\bar{\phi}=const$, $\bar{t}$ coincides with the proper time, that agrees
with the form of the metric (\ref{md2}). It generalizes the corresponding
property of the Kerr metric in Doran coordinates \cite{dor}, \cite{ted11}.

As now $g_{\rho \rho }$ is bounded near the horizon, the proper distance to
the horizon remains finite on the slice $\bar{t}=const$, including the
extremal limit.

\section{Kerr metric}

In the particular case of the Kerr metric, we have in the equatorial plane $%
\theta =\frac{\pi }{2}$ 
\begin{equation}
\frac{d\rho }{dr}=\sqrt{\frac{r^{3}}{r^{3}+ra^{2}+2Ma^{2}}}\text{,}
\end{equation}%
\begin{equation}
N^{2}=\frac{\Delta r^{2}}{(r^{2}+a^{2})^{2}-a^{2}\Delta }=\frac{%
r(r^{2}-2Mr+a^{2})}{r^{3}+ra^{2}+2Ma^{2}}\text{,}
\end{equation}%
\begin{equation}
g_{\phi }=r^{2}+a^{2}+\frac{2M}{r}a^{2}\text{,}
\end{equation}%
\begin{equation}
\omega =\frac{2Ma}{r^{3}+ra^{2}+2Ma^{2}}\text{.}
\end{equation}%
Here, $M\,\ $\ is a black hole mass, $a=\frac{J}{M^{2}}$, $J$ being a black
hole angular momentum. The choice 
\begin{equation}
w=\frac{a}{r^{2}+a^{2}}\sqrt{\frac{2M(r^{2}+a^{2})}{r^{3}+ra^{2}+2Ma^{2}}}
\end{equation}%
satisfies eq. (\ref{hi}), (\ref{w}). Substituting it into (\ref{md2}), we
obtain our metric in the form 
\begin{equation}
ds^{2}=-d\bar{t}^{2}+(\alpha ^{-1}\beta dr+\alpha (d\bar{t}-ad\bar{\phi}%
))^{2}+(r^{2}+a^{2})d\bar{\phi}^{2},
\end{equation}%
\begin{equation}
\beta ^{2}=\frac{2Mr}{r^{2}+a^{2}}\text{, }\alpha ^{2}=\frac{2M}{r}\text{.}
\end{equation}%
that corresponds to eq. (18) of \cite{dor} and eqs. (2), (3) of \cite{ted11}
in which one should put $\theta =\frac{\pi }{2}$.

\section{Circular orbits for near-extremal black holes}

Circular orbits ($r=r_{0}=const$\thinspace $)$ are characterized by the
conditions 
\begin{equation}
Z^{2}(r_{0})=0,  \label{z0}
\end{equation}%
\begin{equation}
\left( \frac{dZ^{2}}{dr}\right) _{r=r_{0}}=0\text{,}  \label{1}
\end{equation}%
where $Z^{2}$ is defined in (\ref{ro}) and we returned from $\rho $ to $r$.

For a black \ hole with arbitrary parameters, the solution of Eqs. (\ref{z0}%
), (\ref{1}) is rather complicated even in the simplest case of the Kerr
metric. For a generic dirty black hole (\ref{met}) it is impossible to find
the solution analytically at all. However, it turns out that if a
nonextremal black hole is close to its extremal state, there are some
general features that enabled us to develop a general approach to the
analysis of such orbits \cite{circ}. If the surface gravity $\kappa =\frac{1%
}{2}(\frac{\partial N^{2}}{\partial \rho })_{\rho =\rho _{H}}$ of a black
hole is small ($\kappa =0$ corresponds to the extremal state), then, it
turns out that for ISCO 
\begin{equation}
N\sim r_{0}-r_{H}=O(\kappa ^{2/3}),  \label{23}
\end{equation}%
and, for the photon and marginally bound near-horizon orbits,%
\begin{equation}
N\sim r_{0}-r_{H}=O(\kappa )\text{.}  \label{k1}
\end{equation}

These relations follow from eqs. (46) and (61) of \cite{circ},
correspondingly.

Thus in the extremal limit $\kappa \rightarrow 0$, the corresponding radius
approaches the horizon, $r_{0}\rightarrow r_{H}$. However, in doing so, the
proper distance within the slice $t=const$ between the horizon and the ISCO 
\begin{equation}
l=O(\ln \frac{1}{\kappa }),
\end{equation}%
as is shown in \cite{circ}. Between the horizon and the marginally bound or
photon orbit 
\begin{equation}
l=O(1)\text{.}
\end{equation}%
These results agree with \cite{72}.

If, instead of $t=const$, we choose the slice $\bar{t}=const,$ the metric is
regular, $g_{\rho \rho }$ is finite in the vicinity of the horizon, so the
proper distance

\begin{equation}
\lim_{\kappa \rightarrow 0}l=0
\end{equation}%
that generalizes observation of Ref. \cite{ted11}.

One reservation is in order. The near-horizon ISCO in the background of
near-extremal black holes exist not always but under some additional
constraints on the metric parameters. See Ref. \cite{circ} and Sec. VII and
IX below. In particular, for the near-extremal Reissner-Nordstr\"{o}m metric
ISCO cannot lie in the vicinity of the horizon (if we restrict ourselves by
geodesic trajectories).

\section{Geometric properties}

The properties of near-horizon obits can be considered from a more general
viewpoint. And, this reveals relation between the issue under discussion and
another issue - namely, so-called Ba\~{n}ados-Silk-West (BSW) effect (see
below). Let us expand the four-velocity of a particle with respect to the
basis that contains two lightlike vectors $l^{\mu }$ and $N^{\mu }$ (so $%
l_{\mu }l^{\mu }=0=N_{\mu }N^{\mu })$ and two spatial vectors. For the case
under discussion, when a particle follows a circular equatorial orbit, $%
r=const$ and also $\theta =const$, so as a matter of fact it is sufficient
to use only two basis vectors $l^{\mu }$ and $N^{\mu }$. We normalize them
in such a way that%
\begin{equation}
l^{\mu }N_{\mu }=-1\text{.}
\end{equation}%
Then,%
\begin{equation}
u^{\mu }=\beta N^{\mu }+\frac{1}{2\alpha }l^{\mu }\text{.}  \label{u}
\end{equation}

It is convenient to choose

\begin{equation}
l^{\mu }=(1,0,0,\omega +\frac{N}{\sqrt{g_{\phi }}}),
\end{equation}

\begin{equation}
N^{\mu }=\frac{1}{2N^{2}}(1,0,0,\omega -\frac{N}{\sqrt{g_{\phi }}})\text{,}
\end{equation}%
$x^{\mu }=(t,r,\theta ,\phi )$. In the horizon limit,%
\begin{equation}
l^{\mu }\rightarrow \xi ^{\mu }+\omega _{H}\eta ^{\mu }\text{,}
\end{equation}%
where $\xi ^{\mu }$ and $\eta ^{\mu }$ are the Killing vectors responsible
for time translation and rotation, respectively.

Equations of motion for a geodesic particle (\ref{t}) - (\ref{phi}) tell us
that on the circular orbit,%
\begin{equation}
X=N\sqrt{\frac{L^{2}}{g_{\phi }}+m^{2}}\text{,}  \label{xc}
\end{equation}%
\begin{equation}
\beta =\frac{1}{m}(X-\frac{LN}{\sqrt{g_{\phi }}})\text{,}
\end{equation}%
\begin{equation}
\frac{1}{\alpha }=\frac{1}{mN^{2}}(X+\frac{NL}{\sqrt{g_{\phi }}})\text{.}
\end{equation}

Thus, for small $N$, we have from (\ref{xc}) that%
\begin{equation}
\beta =pN=O(N)\text{,}  \label{be}
\end{equation}%
\begin{equation}
\alpha =qN=O(N)\,\text{,}  \label{al}
\end{equation}%
where the coefficients are equal to%
\begin{equation}
p=\frac{1}{m}(\sqrt{\frac{L^{2}}{g_{\phi }}+m^{2}}-\frac{N}{\sqrt{g_{\phi }}}%
)_{H}=O(1)\text{,}
\end{equation}%
\begin{equation}
q^{-1}=\frac{1}{m}(\sqrt{\frac{L^{2}}{g_{\phi }}+m^{2}}+\frac{N}{\sqrt{%
g_{\phi }}})_{H}=O(1).
\end{equation}%
Also, it follows from (\ref{t}) - (\ref{phi}) that%
\begin{equation}
\frac{d\phi }{dt}=\frac{u^{\phi }}{u^{t}}=\omega _{H}+O(N)\text{,}
\end{equation}%
\begin{equation}
\frac{l^{\phi }}{l^{t}}=\omega _{H}+O(N)\text{.}
\end{equation}%
In this sense, the orbit does indeed become asymptotically parallel to the
horizon generator, for which $\frac{d\phi }{dt}=\omega _{H}$. However,
although the coefficient at $l^{\mu }$ (which, in turn, approaches the
horizon generator) is much larger than that at $N^{\mu }$, both terms in (%
\ref{u}) give comparable contribution to the norm of $u^{\mu }$ \cite{gp}
because of the property $l_{\mu }N^{\mu }=0$.

\section{Extremal black hole and fake trajectory on horizon}

In the previous section, we discussed the limit $\kappa \rightarrow 0$. What
happens if $\kappa =0$ from the very beginning? Formally, one is tempted to
put $\kappa =0$ in (\ref{23}), (\ref{k1}).\ However, the process of
derivation in \cite{circ} essentially implied that although $\kappa $ is
small, $\kappa \neq 0.$ For $\kappa =0$, the asymptotic expansion of the
metric coefficient has another form as compared to the nonextremal case, so
it is necessary to proceed anew. According to general properties, for the
extremal black holes the expansion of the metric coefficient $\omega $ reads 
\cite{t}

\begin{equation}
\omega =\omega _{H}-B_{1}N+B_{2}N^{2}+O(N^{3})\text{,}  \label{omb}
\end{equation}%
whence one obtains from (\ref{X}) that%
\begin{equation}
X=X_{H}+(B_{1}N-B_{2}N^{2})L+O(N^{3})\text{,}  \label{xh}
\end{equation}%
where $B_{1}$ and $B_{2}$ are model-dependent coefficients.

If $X_{H}=0$, it seems that there exists an exact solution eqs. (\ref{z0}), (%
\ref{1}) that reads $N=0$, $X=0$. It \textit{would seem }that it describes
the trajectory that lies within the horizon. However, it contradicts the
fact that the time-like trajectory cannot lie on the null surface.

In Introduction in \cite{m}, the attempt was made to assign a meaning to
such orbits on the extremal horizon. It is based on the results of \cite%
{ted11}, where it was revealed that the proper distance to the horizon
depends essentially on a slice (see also Sec. II - IV above). Then, an orbit
can in a sense lie on the horizon, if the slice $\bar{t}=const$ of the Doran
coordinate system \cite{dor} is chosen. However, for the situation discussed
in \cite{ted11}, black holes are near-extremal but not exactly extremal,
there is a parameter $\kappa $ that can be as small as one likes but
nonzero. The proper distance between the circular orbit and the horizon
remains finite for any $\kappa $ on both types of slices ($t=const$ or $\bar{%
t}=const$). Only asymptotically, in the limit $\kappa \rightarrow 0$, the
proper distance to the horizon on the slice $\bar{t}=const$ approaches zero,
so the claim "the circular orbit lies on the horizon" is to be understood in
the asymptotic sense. But now, for extremal black holes, $\kappa =0$ exactly
from the very beginning, so the reference to the approach of \cite{ted11}
does not save the matter. Here, the situation should be considered anew.

Actually, we have in (\ref{ro}) with $\frac{d\rho }{d\tau }=\frac{dr}{d\tau }%
=0$ the uncertainty of the type \thinspace $\frac{0}{0}$. To resolve this
uncertainty, the original coordinates are insufficient since they become
degenerate on the horizon. Instead, we use the barred coordinates in which
the metric takes the form (\ref{md}).

Below, we generalize approach of Sec.III C of \cite{circkerr}, where this
issue was considered for the Kerr metric. For the circular orbit, $\frac{%
d\rho }{d\tau }=0$. On the horizon, $\bar{N}=0$, $g_{\bar{t}\bar{t}}=g_{\phi
}\omega _{H}^{2}$. After substitution into (\ref{E}), (\ref{L}), we see that
in this case%
\begin{equation}
\mathcal{E}-\omega _{H}\mathcal{L}=0\text{.}  \label{crit}
\end{equation}%
Then, direct calculation gives us%
\begin{equation}
u_{\mu }u^{\mu }=\mathcal{E}^{2}\frac{1}{g_{\phi }\omega _{H}^{2}}\text{.}
\end{equation}

As $u_{\mu }u^{\mu }=-1$ for time-like curves and $u_{\mu }u^{\mu }=0$ for
lightlike ones, we conclude that for trajectories of physical particles the
only possibility is to $\mathcal{E}=0=\mathcal{L}$, the curve must be
light-like. Thus the time-like on-horizon orbit is fake.

We also infer from (\ref{E}) and (\ref{L}) that%
\begin{equation}
\frac{d\bar{\phi}}{dt}=\omega _{H}\text{.}
\end{equation}%
Actually, it means that for massless particles, the trajectory in question
coincides with the horizon generator.

\section{Circular orbits for near-critical particles}

In what follows, we use terminology borrowed from works on studies of the
BSW effect. We call a particle critical if $X_{H}=0$ and usual if $X_{H}>0$
is generic. If $X_{H}\neq 0$ but is small, we call a particle near-critical.
It follows from (\ref{crit}) that a particle whose trajectory lies on the
horizon is necessarily critical. Such a trajectory can be realized for
photon and is forbidden for massive particles. Meanwhile, although a
trajectory that lies exactly on the horizon is impossible for massive
particles, near-horizon circular orbits are allowed, if corresponding
particles are near-critical (not exactly critical), as we will see it below.

It is convenient to use $N$ as a radial coordinate. Then, analogue of eq. (%
\ref{1}) reads%
\begin{equation}
(\frac{dZ^{2}}{dN})_{N=N_{0}}=0\text{,}  \label{zn}
\end{equation}%
where $N_{0}$ corresponds to the circular orbit. Then, it follows from (\ref%
{z0}) and (\ref{zn}) that%
\begin{equation}
X_{0}=N_{0}\sqrt{Y_{0}}\text{,}  \label{xn}
\end{equation}%
\begin{equation}
\frac{1}{2}(\frac{dZ^{2}}{dN})_{0}=N_{0}(B\sqrt{Y}L-Y-\frac{N}{2}\frac{dY}{dN%
})_{0}\text{,}  \label{1d}
\end{equation}%
where%
\begin{equation}
Y=\frac{L^{2}}{g}+m^{2}\text{,}  \label{Y}
\end{equation}%
\begin{equation}
B=-\frac{d\omega }{dN}\text{,}
\end{equation}%
subscript "0" means that the corresponding quantities are calculated on the
circular orbit.

To avoid fake orbits with $N_{0}=0$, we are interested in the solution with $%
N_{0}\neq 0$, so we have%
\begin{equation}
(B\sqrt{Y}-Y-\frac{N}{2}\frac{dY}{dN})_{0}=0\text{.}  \label{b}
\end{equation}

Thus there are two equations (\ref{xn}) and (\ref{b}) for three unknowns $%
X_{0}$, $N_{0}$, $L$. We can fix, say, $N_{0}$ and, in principle, calculate $%
X_{0}$ and $L_{0}$ (or $E$ and $L_{0})$. Eq. (\ref{b}) is exact.

If we are interested in just near-horizon orbits, we must require $%
N_{0}\approx 0$ and solve the equations iteratively in the form of the
Taylor expansion%
\begin{equation}
L=L_{0}+L_{1}N_{0}+L_{2}N_{0}^{2}+...  \label{ln}
\end{equation}

In the main approximation in which corrections due to small $N_{0}$ are
discarded, we obtain%
\begin{equation}
B_{1}L_{0}=\sqrt{Y_{H}}=\sqrt{\frac{L_{0}^{2}}{g_{H}}+m^{2}}\text{,}
\label{s}
\end{equation}%
where $B_{1}$ is the coefficient entering the expansion (\ref{omb}). Also,
in the same approximation one can take $X_{H}\approx 0$ according to (\ref%
{xn}), whence%
\begin{equation}
L_{0}=\frac{E}{\omega _{H}}\text{.}  \label{le}
\end{equation}

It follows from (\ref{s}), (\ref{le}) that%
\begin{equation}
L_{0}=\frac{m}{\sqrt{B_{1}^{2}-\frac{1}{g_{H}}}}\text{, }E=\frac{m\omega _{H}%
}{\sqrt{B_{1}^{2}-\frac{1}{g_{H}}}}\text{,}  \label{lme}
\end{equation}%
where we assume that $m\neq 0$ (for the case $m=0$, see below). The formulas
(\ref{lme}) imply that 
\begin{equation}
B_{1}\sqrt{g_{H}}>1\text{.}  \label{bg}
\end{equation}

If (\ref{s}), (\ref{bg}) are violated, there are no circular near-horizon
orbits for massive particles. Eq. (\ref{s}) constraints black hole
parameters, for example the angular momentum of a black hole. In particular,
for the Kerr-Newman black hole this leads to some distinct values of the
black hole angular momentum \cite{m} (see also Sec. VII and IX below). In
particular, for the Reissner-Nordstr\"{o}m metric $\omega =0=B_{1}$, eq. (%
\ref{bg}) is violated and this means that the near-horizon ISCO does not
exist in this case.

For different types of orbits eq. (\ref{s}) or (\ref{lme}) leads to
different relations.

\subsection{Photon orbit}

Putting $m=0$ in (\ref{s}), we obtain%
\begin{equation}
B_{1}^{2}=\frac{1}{g_{H}}\text{.}  \label{phg}
\end{equation}

\subsection{Marginally bound orbit}

Putting $E=m,$ in the zero approximation we obtain from (\ref{z0}) that%
\begin{equation}
L_{0}=\frac{m}{\omega _{H}}\text{.}  \label{mbg}
\end{equation}%
Then, (\ref{s}) gives us%
\begin{equation}
B_{1}^{2}=\frac{1}{g_{H}}+\omega _{H}^{2}\text{.}  \label{mbge}
\end{equation}

\subsection{ISCO}

By definition, the condition%
\begin{equation}
\left( \frac{d^{2}Z^{2}}{dz^{2}}\right) _{0}=0  \label{2d}
\end{equation}%
should be satisfied for this orbit in addition to (\ref{z0}) and (\ref{zn})
since eq. (\ref{2d}) gives the boundary between stable and unstable orbits.
In the main approximation, $\frac{1}{2}\left( \frac{d^{2}Z^{2}}{dz^{2}}%
\right) _{0}\approx N_{0}L^{2}(\frac{1}{g^{2}}\frac{dg}{dN}-2B_{2}B_{1})_{0}$%
, so we have%
\begin{equation}
S\equiv (\frac{1}{g^{2}}\frac{dg}{dN}-2B_{2}B_{1})_{H}=0\text{,}  \label{z02}
\end{equation}%
where we neglected in (\ref{z02})\ the difference between the quantities
calculated on the circular orbit and on the horizon. In addition to (\ref%
{z02}), eq. (\ref{lme}) should be satisfied.

\subsection{Estimate of $X_{H}$}

It is instructive to find $X_{H}$ for all types of orbits under discussion.
It follows from (\ref{xh}), (\ref{xn}) and (\ref{Y}) that%
\begin{equation}
X_{H}=N_{0}(\sqrt{Y}-B_{1}L)+O(N_{0}^{2})\text{.}
\end{equation}%
The terms of the order $O(N_{0})$ mutually cancel, so%
\begin{equation}
X_{H}=O(N_{0}^{2})\text{.}  \label{xoh}
\end{equation}%
Thus the particle that moves on the circular orbit near the extremal black
hole turns out to be not only near-critical but should have anomalously
small $X_{0}$ to keep following such an orbit.

\subsection{Extremal Kerr-Newman black hole}

For the equatorial orbit ($\theta =\frac{\pi }{2}$) in the extremal
Kerr-Newman background, one can calculate the relevant quantities (using,
say, the Boyer-Lindquiste coordinates) and find

\begin{equation}
g_{H}=\frac{(M^{2}+a^{2})^{2}}{M^{2}}\text{,}  \label{gh}
\end{equation}%
\begin{equation}
\left( \frac{dg}{dN}\right) _{H}=2\frac{(M^{2}-a^{2})}{M^{4}}%
(M^{2}+a^{2})^{2}\text{,}
\end{equation}

\begin{equation}
B_{1}=\frac{2a}{M^{2}+a^{2}}\text{,}  \label{b1m}
\end{equation}%
\begin{equation}
B_{2}=\frac{a^{3}}{M^{2}(M^{2}+a^{2})}\text{,}
\end{equation}%
\begin{equation}
\omega _{H}=\frac{a}{M^{2}+a^{2}}\text{,}  \label{okn}
\end{equation}%
\begin{equation}
S=\frac{2(M^{2}-2a^{2})}{M^{2}(M^{2}+a^{2})}\text{.}  \label{S}
\end{equation}%
Then, one can find from eqs. (\ref{s}), (\ref{le}), (\ref{S}) that $a=a_{0}$%
, where 
\begin{equation}
\frac{a_{0}}{M}=\frac{1}{2},\frac{1}{\sqrt{3}},\frac{1}{\sqrt{2}}  \label{a0}
\end{equation}%
$\ $\ for the photon orbit, marginally bound one and ISCO, respectively.
These values agree with \cite{bal}, \cite{m}.

\section{Dynamics of critical particles}

In the previous section we saw that for a circular orbit to exist in the
immediate vicinity of the horizon, the relations (\ref{s}), (\ref{ln})
should be satisfied. It \textit{would seem} that, as these relations imply $%
N_{0}=0$, the corresponding orbit lies exactly on the horizon. However, we
already know that such an orbit is fake for massive particles. We may look
for the solution for circular orbit in a series with respect to $N_{0}$.
Then, apart from (\ref{ln}), the similar expansion should go, say, for the
parameter $a$ (if, for definiteness, we consider the Kerr-New-Newman metric):%
\begin{equation}
a_{c}=a_{0}+a_{1}N_{0}+O(N_{0}^{2})\text{,}  \label{ac}
\end{equation}%
where $a_{0}$ is the aforementioned value, different for different kinds of
orbits, $a_{c}$ corresponds to the circular orbit.

Let, say, $a_{1}<0$ but a black hole has $a>a_{0}$. Then, the contradiction
with (\ref{ac}) tells us that for $a>a_{0}$ the circular orbits do not exist
near the horizon at all. For example, for the Kerr-Newman metric there is an
exact solution describing the circular photon orbit \cite{bal} $r_{0}=2M-2a$
that can be rewritten as%
\begin{equation}
a=\frac{M}{2}-\frac{1}{2}(r_{0}-M)<\frac{M}{2}
\end{equation}%
for any orbit above the horizon. If $a>\frac{M}{2}$, such a solution ceases
to exist.

Formally, then it follows from (\ref{1d}) that the only solution is $N_{0}=0$
(the orbit exactly on the horizon). However, we reject this case since for
massive particles it is impossible and for massless ones it is already
described above. Thus we are led to the conclusion that, in the absence of a
circular orbits, a particle should move. We assume that it moves towards a
black hole.

To probe dynamics in this situation, it is instructive to select particles
which are exactly critical ($X_{H}=0)$ since it is this value that was a
"candidate" for a circular orbit on the horizon.

It is convenient to expand $Z^{2}$ in powers of $N$:%
\begin{equation}
Z^{2}=z_{2}N^{2}+z_{3}N^{3}+...  \label{zx}
\end{equation}

Here, the coefficients%
\begin{equation}
z_{2}=(B_{1}^{2}-\frac{1}{g_{H}})L^{2}-m^{2}
\end{equation}%
\begin{equation}
z_{3}=[\frac{1}{g_{H}^{2}}(\frac{dg}{dN})_{H}-2B_{1}B_{2}]L^{2}\text{,}
\end{equation}%
correspond just to (\ref{s}), (\ref{z02}) but now they are, on general, do
not vanish.

It is also convenient to write%
\begin{equation}
N=F(\rho )(\rho -\rho _{H})=H(r)(r-r_{H})\text{, }F(\rho _{H})\equiv
F_{1}\neq 0\text{, }H(r_{H})=H_{1}\neq 0\text{.}
\end{equation}

Then, for $z_{2}>0$ we obtain from (\ref{ro})%
\begin{equation}
\frac{dN}{d\tau }\approx -F_{1}N\sqrt{z_{2}}\text{,}
\end{equation}%
\begin{equation}
r-r_{H}\approx r_{1}\exp (-F\sqrt{z_{2}}\tau )\text{,}
\end{equation}%
where $r_{1}$ is another constant.

Let now $z_{2}=0$, $z_{3}>0$. In a similar way, we find from (\ref{ro}) that%
\begin{equation}
\frac{1}{F}\frac{dN}{d\tau }\approx -\sqrt{z_{3}}N^{3/2}\text{,}
\end{equation}

\begin{equation}
r-r_{H}\approx \frac{4}{F_{1}^{2}\tau H_{1}^{2}z_{3}}.
\end{equation}

In general, if $X_{H}=0$ and  $z_{2}(r_{+})=z_{3}(r_{+})=...z_{k}=0$, $%
z_{k+1}>0$,%
\begin{equation}
r-r_{H}\sim \lambda ^{-\frac{2}{k-1}}\text{,}
\end{equation}%
where $\lambda $ is the affine parameter (proper time for a massive
particle).

\section{Circular orbits in vicinity of near-extremal black holes}

We saw that the circular orbits exist for selected values of black hole
parameters only (say, the discrete values of the angular momentum $a=a_{0}$%
). In the main approximation, these values do not depend on the posiiton of
the near-horizon orbit, in the next approximation $a_{0}$ acquires small
corrections of the order $N_{0}$. This is in sharp contrast with the
situation for near-extremal black holes, when the surface gravity $\kappa $
is small but nonzero. It is shown in \cite{circ} that for such black holes
circular orbits almost always exist, under rather weak restriction that
black hole parameters obey some inequalities (see below). For example, for
the case the Kerr-Newman metric it means that $M^{2}$ is to be close to $%
a^{2}+Q^{2}$ (where $Q$ is the black hole electric charge) but the ratio $%
\frac{a}{M}$ is arbitrary in some finite interval. Thus, the situation for
extremal black holes in the aspects under discussion cannot be considered as
a limit $\kappa \rightarrow 0$ of that for near-extremal black holes.
Actually, we have two distinct situations.

1) Near-extremal black holes. Here, black hole parameters are free ones that
change contiuously in some interval. The radius of the circular orbit $r_{0}$
near the horizon is not arbitrary but is defined by black hole parameteres
from equilibrium conditions \cite{circ}.

2) Extremal black holes. Here, black hole parameters are fixed but the
quantity $N_{0}$ that characterizes the location of the orbit is a small
free parameter.

In a sense, both cases are complimentary to each other.

To gain further insight, how it happens, let us consider briefly the case of
near-extremal black holes (for more details one can consult ref. \cite{circ}%
). Near the horizon,%
\begin{equation}
N^{2}\approx 2\kappa x+Dx^{2}\text{,}  \label{nx}
\end{equation}%
\begin{equation}
\omega \approx \omega _{H}-b_{1}x\text{,}  \label{ob}
\end{equation}%
where $x=\rho -\rho _{H}$, $\kappa $ is a small but nonzero parameter, $D~$%
and $b_{1}$ are constants. Then, eq. (\ref{z0}), (\ref{1}) give us, in the
main approximation (say, for the marginally bound orbit)%
\begin{equation}
\sqrt{2\kappa x+Dx^{2}}b_{1}L=(\kappa +Dx)\sqrt{Y_{H}}\text{.}  \label{yd}
\end{equation}

Taking into account (\ref{omb}), (\ref{ob}), (\ref{nx}), we see that $%
b_{1}=B_{1}\sqrt{D}$, whence%
\begin{equation}
B_{1}L=\sqrt{Y_{H}}\frac{\kappa +Dx}{\sqrt{2\kappa Dx+D^{2}x^{2}}}.
\label{dk}
\end{equation}

If $\kappa \neq 0$, it follows from (\ref{dk}) that%
\begin{equation}
B_{1}L>\sqrt{Y_{H}}\text{.}  \label{by}
\end{equation}%
In view (\ref{Y}), it is seen that for the existence of the circular orbit, (%
\ref{bg}) should be satisfied. It is worth noting that eq. (\ref{by})
corresponds to eq. (44) of \cite{circ} in which one should require
positivity of the quantity $L_{0}^{2}$.

We can make the substitution $x=\kappa D\alpha $ and obtain the equation
with respect to $\alpha $,%
\begin{equation}
\sqrt{\alpha ^{2}+2\alpha }B_{1}L_{0}=(1+\alpha )\sqrt{Y_{H}}.  \label{ay}
\end{equation}%
For the marginally bound orbit, we should insert $L_{0}=\frac{m}{\omega _{H}}
$ in (\ref{ay}). Its solution gives us $\alpha $ as a quite definite
function of black hole parameters $B_{1}$, $\omega _{H}$, $g_{H},$ so the
location of the orbit near the horizon is fixed. This is done for any small $%
\kappa $.

In a similar way, for the photon orbit we substitute $m=0$ into (\ref{ay}).
Then, we have another equation for $\alpha $:%
\begin{equation}
\sqrt{\alpha ^{2}+2\alpha }B_{1}=\frac{1+\alpha }{\sqrt{g_{H}}}\text{.}
\label{phne}
\end{equation}

By contrast, if $\kappa =0$ exactly, we obtain from (\ref{yd}) the equation $%
x(B_{1}L-\sqrt{Y_{H}})=0$. As the orbit does not lie on the horizon, $x\neq
0 $, and we infer from it that the condition (\ref{s}) is to be satisfied.

Thus we see that a nontrivial play of small quantities $x$ and $\kappa $
takes place in such a way that the same eqs. (\ref{z0}), (\ref{1}) constrain
the different entities for near-extremal and extremal black holes. In the
first case, it is the location of the circular orbit, in the second one it
is the restriction on black hole parameters.

Let us consider now the situation for the ISCO. Then, direct calculations of
the second derivative in eq. (\ref{2d}) shows that eqs. (\ref{s}), (\ref{lme}%
) are indeed satisfied. (This corresponds to eq. 37 of \cite{circ}.) Now $E$
and $L_{0}$ themselves can be found from these equations and give rise to
eq. (\ref{lme}). Eq. (\ref{1}) for the ISCO requires more terms in the
expansion then in (\ref{yd}) and leads to (\ref{23}), as is shown in \cite%
{circ} in agreement with the previous results on the Kerr metric \cite%
{circkerr}.

For all thee types of orbits, inequality (\ref{bg})\ on black hole
parameters should be satisfied but, unlike the extremal case, it does not
select any discrete values of them. Say, for the Kerr-Newman metric one can
only infer from (\ref{bg}) that $\frac{a}{M}>\frac{1}{2}$.

Let us summarize which equations govern the behavior of which orbits. For
near-extremal black holes, these are eq. (\ref{ay}) with $E=m$ for the
marginally bound orbit or eq. (\ref{phne}) for the photon orbit. For the
ISCO, this is eq. (\ref{s}) and one more equation that is the consequence of
(\ref{z0}) and (\ref{1}). It looks more complicated and can be found in eqs.
(36), (39)\ - (41) of \cite{circ}. In all three cases one finds $x_{0}$ as
the function of the black hole parameters.

For pure extremal black holes the equaitons under discussion are (\ref{s})
with $E=m$ (the marginally bound orbit) or (\ref{s}) with $m=0$ (the photon
orbit). For the ISCO these are (\ref{s}) and eq. (\ref{z02}) (where for the
Kerr-Newman metric\thinspace $\ S$ should be taken from (\ref{S})). The
aforementioned equations give constraints on black hole parameters. The
radius of the orbit is arbitrary but in the near-horizon region is
restricted by the condition $N_{0}\ll 1$.

\section{Velocity on circular near-horizon orbits}

It was observed in \cite{72} that in the extremal limit of the Kerr metric
the velocity $V$ measured by a locally nonrotating observer is not equal to
1, as one could na\"{\i}vely expect. For the ISCO, it was found that $V=%
\frac{1}{2}$. For the marginally bound orbit, it turned out that $V=\frac{1}{%
\sqrt{2}}$. Now, we generalize these results, obtain the similar ones for
pure extremal black holes and compare the both.

If a particle moves in the background of the stationary metric (\ref{met}),
the following relation holds: 
\begin{equation}
X=\frac{mN}{\sqrt{1-V^{2}}},  \label{mn}
\end{equation}%
see eq. (15) of \cite{k}. Taking into account also eq. (\ref{xn}) for the
circular orbit, one obtains%
\begin{equation}
V=\frac{L_{0}}{\sqrt{L_{0}^{2}+m^{2}g_{H}}}\text{.}
\end{equation}

\subsection{Near-extremal black holes}

Different orbits should be considered separately.

\subsubsection{Marginally bound orbit}

Now, taking into account (\ref{mbg}), we obtain%
\begin{equation}
V=\frac{1}{\sqrt{1+\omega _{H}^{2}g_{H}}}\text{.}  \label{vmb}
\end{equation}%
For the extremal Kerr-Newman black hole, it is seen from (\ref{gh}) and (\ref%
{okn}) that $\omega _{H}^{2}g_{H}=\frac{a^{2}}{M^{2}}$, so we obtain%
\begin{equation}
V=\frac{M}{\sqrt{M^{2}+a^{2}}}\text{.}  \label{vma}
\end{equation}

In the extremal Kerr case, $a=M\,$, so $V=\frac{1}{\sqrt{2}}$ in agreement
with \cite{72}.

\subsubsection{ISCO}

In this case, taking into account (\ref{lme}), one obtains%
\begin{equation}
V=\frac{1}{B_{1}\sqrt{g_{H}}}\text{.}  \label{visco}
\end{equation}%
It is implied that (\ref{bg}) is satisfied, so $V<1$.

For the extremal Kerr-Newman metric, it folows from (\ref{gh}) and (\ref{b1m}%
) that $B_{1}\sqrt{g_{H}}=2\frac{a}{M}$, so 
\begin{equation}
V=\frac{M}{2a}\text{,}  \label{visca}
\end{equation}%
where now $a>\frac{M}{2}$ for the existence of the orbit under discussion.
In the extremal Kerr case, $a=M$ and we obtain $V=\frac{1}{2}$ in agreement
with \cite{72}.

\subsubsection{Photon circular orbit}

Now, by analogy with the massive case, we can introduce $X=\nu _{0}-\omega L$%
, where $\nu _{0}$ is the conserved quantity having the meaning of the
frequency measured at infinity, $\nu $ is the locally measured frequency 
\cite{k}. Instead of (\ref{mn}), now%
\begin{equation}
\nu _{0}-\omega L=\nu N\text{,}
\end{equation}%
(see eq. 40 of \cite{k}). In this case, instead of velocity, it makes sense
to speak about the effective gamma-factor $\gamma =\frac{\nu }{\nu _{0}}$.
Using (\ref{xn}) with $m=0$, one obtains for the near-critical particle with 
$\nu _{0}\approx \omega _{H}L$ that%
\begin{equation}
\gamma =\frac{1}{\omega _{H}\sqrt{g_{H}}}\text{.}  \label{gap}
\end{equation}%
For the Kerr-Newman case, it is seen from (\ref{gh}), (\ref{okn}) that $%
\omega _{H}\sqrt{g_{H}}=\frac{a}{M}$. After substitution into (\ref{gap}),
one finds%
\begin{equation}
\gamma =\frac{M}{a}.  \label{gam}
\end{equation}

In the extremal Kerr case $\gamma =1$.

\subsection{Extremal black holes}

The previous formulas for the velocity are valid but with the additional
constraint that now (\ref{s}) should be satisfied. Correspondingly, for the
Kerr-Newman case $a$ is no longer a free parameter but equal to $a_{0}$,
where $a_{0}$ should be taken from eq. (\ref{a0}) and substituted into (\ref%
{vma}) or (\ref{gam}). Below, we list the values of the velocity on
near-horizon circular orbits for the Ker-Newman metric.

\subsubsection{Marginally bound orbit}

By substitution $a=\frac{1}{\sqrt{3}}$ into (\ref{vma}), we get

\begin{equation}
V=\frac{\sqrt{3}}{2}
\end{equation}

\subsubsection{ISCO}

Now, $a=\frac{1}{\sqrt{2}}$, and we find from (\ref{visca}) that

\begin{equation}
V=\frac{1}{\sqrt{2}}\text{.}
\end{equation}%
Thus, the same value $V=\frac{1}{\sqrt{2}}$ is obtained for the marginally
bound orbit in the near-extremal case and for the ISCO in the pure extremal
one.

\subsubsection{Photon orbit}

For this orbit, $a=\frac{1}{2}$, so (\ref{gam}) gives us

\begin{equation}
\gamma =2\text{.}
\end{equation}

Thus we see that the values of the velocity or Lorentz factor on the
near-horizon orbit cannot be obtained as a limit $\kappa \rightarrow 0$ of
these values on orbits in the vicinity of near-extremal black holes.

\subsection{Radially moving critical particle}

For comparison, we also consider the case when a particle is exactly
critical, $E=\omega _{H}L$. The particle's orbit is not circular, the
particle approaches the horizon. Then, using (\ref{xh}) instead of (\ref{xn}%
), one finds from (\ref{mn}) that%
\begin{equation}
V=\sqrt{1-\frac{\omega _{H}^{2}m^{2}}{B_{1}^{2}E^{2}}}\text{.}
\end{equation}%
For the Kerr-Newman case, $\frac{B_{1}}{\omega _{H}}=2$, so%
\begin{equation}
V=\sqrt{1-\frac{m^{2}}{4E^{2}}}\text{.}
\end{equation}%
If a particle falls from infinity with the zero initial velocity, $m=E$.
Then, $V=\frac{\sqrt{3}}{2}$.

\section{General features of circular and would-be circular orbits in the
near-horizon region}

Thus there are two typical kinds of circular or almost circular orbits in
the immediate vicinity of the horizon. (i) There exist true circular orbits
that require the particle to be very close to the critical state (\ref{xoh})
but not coincide with the critical one nonetheless. For a fixed black hole
metric, this is possible for some special values of parameters only. Say,
for the Kerr-Newman black hole the parameter $\frac{a}{M}=\frac{1}{\sqrt{2}}$%
, $\frac{1}{\sqrt{3}}$, $\frac{1}{2}$ (plus corrections $O(N_{0})$) for the
ISCO, marginally bound and photon orbits respectively. (ii) One can specify $%
X_{H}=0$ (or take $X_{H}=o(N_{0}^{2})$ )$.$Then, circular orbits do not
exist at all in the near-horizon region. Instead, a particle inspirals
approaching the horizon asymptotically that takes an infinitely long proper
time \cite{ted}, \cite{gp}, \cite{prd}. The rate is either exponential or
power-like depending on a black hole parameters.

In the case of the extremal Kerr-Newman metric the concrete characteristic
values of $\frac{a}{M}$ found above coincide with those in \cite{m} the
first approximation. However, interpretation is qualitatively different. In
case (i) the true circular orbits of massive particles are absent from the
horizon $N=0$. In case (ii) these values determine the rate with which the
particle approaches the horizon.

\section{Relation to the Ba\~{n}ados-Silk-West effect and the kinematic
censorship}

In the near-horizon region, $N$ is small, so on the circular orbit $X$ is
also small according to (\ref{xc}). This is realized for small surface
gravity $\kappa $ (\ref{23}), (\ref{1}), i.e. for near-extremal black holes.
Meanwhile, such orbits play an essential role in the so-called the Ba\~{n}%
ados-Silk-West (BSW) effect \cite{ban}. Namely, if two particles collide
near a black hole, the energy centre of mass $E_{c.m.}$ grows unbound,
provided one of particles is critical or near-critical. For the orbits under
discussion, according to (\ref{xn}), $X$ has the order $N$, so a
corresponding particle is near-critical. Let us consider in more detail, how
the properties of the orbit are related to the BSW effect.

Let two particles with masses $m_{1}$ and $m_{2}$ collide. By definition,
the energy in the centre of mass%
\begin{equation}
E_{c.m.}^{2}=-P_{\mu }P^{\mu }=m_{1}^{2}+m_{2}^{2}+2m_{1}m_{2}\gamma \text{,}
\label{cm}
\end{equation}%
where $P^{\mu }=p_{1}^{\mu }+$ $p_{2}^{\mu }$ is their total momentum, $%
p_{1.2}=m_{1,2}$u$_{1,2}^{\mu }$, the Lorentz factor of relative motion 
\begin{equation}
\gamma =-u_{1\mu }u_{2}^{\mu }.  \label{ga}
\end{equation}

Using the expansion (\ref{u}), one obtains%
\begin{equation}
\gamma =\frac{1}{2}(\frac{\beta _{1}}{\alpha _{2}}+\frac{\beta _{2}}{\alpha
_{1}})\text{.}  \label{gab}
\end{equation}%
If a particle 1 orbits a black hole near the horizon while particle 2 is
usual, it follows from (\ref{be}), (\ref{al}) that%
\begin{equation}
\gamma \approx \frac{1}{2N}\frac{\beta _{2}}{q_{1}}
\end{equation}%
becomes unbound. The above formulas can be viewed as modification of
approach of \cite{cqg}, where it was applied to radial motion, to the case
of orbital motion.

On one hand, the proximity of the four-velocity to the horizon generator is
due to the fact that the coefficient at $l^{\mu }$ in (\ref{u}) $\ \alpha
_{1}^{-1}=O(N^{-1})$ is much larger than the coefficient at $N^{\mu }$ equal
to $\beta _{1}=O(N)$ . From the other hand, the ratio $\frac{\beta _{2}}{%
\alpha _{1}}$ enters the expression for the Lorentz factor, so the same
parameter $N^{-1}$ controls both phenomena.

In this context, it is worth reminding that collision between one critical
and one usual particle leads to unbound $E_{c.m.}=O(N^{-1/2})$ \cite{prd}.
Had the collision been possible on the horizon exactly ($N=0$) we would have
obtained infinite $E_{c.m.}$ And, as a proper time required to reach the
horizon is finite for a usual particle, these divergences would have been,
at least in principle, observable. This would be unphysical: in any event
the energy that can be obtained in any frame cannot be infinite (it can be
called "principle of kinematic censorship"). Fortunately, the trajectory of
a massive particle on the horizon is impossible, so the experiment under
discussion is impossible as well.

However, there is one more potentially dangerous scenario. It is realized
when particle 1 is massless that needs a separate treatment since in this
case the orbit of particle 1 is not fake. In this case, eqs. (\ref{u}), (\ref%
{gab}) are not applicable directly since the particle is characterized by a
light-like wave vector $k^{\mu }$ (or $p^{\mu }=$%
%TCIMACRO{\U{127}}%
%BeginExpansion
h{\hskip-.2em}\llap{\protect\rule[1.1ex]{.325em}{.1ex}}{\hskip.2em}%
%EndExpansion
$k^{\mu }$, where 
%TCIMACRO{\U{127} }%
%BeginExpansion
h{\hskip-.2em}\llap{\protect\rule[1.1ex]{.325em}{.1ex}}{\hskip.2em}
%EndExpansion
is the Planck constant) instead of the time-like vector $u^{\mu }$. Formulas
(\ref{cm}) and (\ref{ga}) need some modification. Then, 
\begin{equation}
E_{c.m.}^{2}=m^{2}+2m\gamma \text{,}
\end{equation}%
\begin{equation}
\gamma =-u_{\mu }p^{\mu }\text{.}  \label{up}
\end{equation}%
According to general rules, if a photon 1 is critical and the massive
particle 2 is usual, $\gamma $ grows unbound for collisions near the horizon 
\cite{k}.

Let a usual massive particle 2 cross the extremal horizon where it collides
with the photon (particle 1) that follows the horizon generator exactly. Is
the quantity $E_{c.m.}$ finite or infinite?

To answer this question, we need to evaluate the scalar product (\ref{up}).
It is conveneint to use the coordinate system in which the metric has the
form (\ref{md}). The terms with $\mu =\bar{t}$ and $\mu =\bar{\phi}$ do not
contribute since for the photon under consideration $\bar{E}=0=\bar{L}$ as
is explained in\ Sec. VI. Taking also into account that for photon moving
along the horizon $\rho =const$, we obtain that%
\begin{equation}
-p_{\mu }u^{\mu }=(g_{\phi }\omega ^{2}-1)_{H}\left( \frac{d\rho }{d\tau }%
\right) _{2}\left( \frac{d\bar{t}}{d\lambda }\right) _{1}\text{,}
\end{equation}%
$\lambda $ is an affine parameter for the photon trajectory, we took into
account (\ref{md}) and (\ref{ww}).

But $\left( \frac{d\rho }{d\tau }\right) _{2}$ is finite as it follows from (%
\ref{ro}). It is easy to understand that $\left( \frac{d\bar{t}}{d\lambda }%
\right) _{1}$ should be finite as well since this quantity is defined in any
point of a horizon for any value of an affine parameter $\lambda $ and the
coordinates (\ref{md}) are regular on the horizon. Therefore, a particle
cannot have infinite $\left( \frac{d\bar{t}}{d\lambda }\right) _{1}$ in all
points of its trajectory. As a result, the Lorentz factor of relative motion 
$\gamma $ and $E_{c.m.}$ are also finite. Therefore, the kinematic
censorship is preserved.

According to general picture described in detail in \cite{prd}, \cite{k},
collision between one usual and one critical particle leads to the BSW
effect.\ However, in all cases discussed in the aforementioned references,
there exists a small but nonzero parameter that controls the process. For a
particle slowly moving in a radial direction and approaching the extremal
horizon this is $\tau ^{-1}$ since the proper time $\tau $ required for the
critical particle to reach the horizon is infinite \cite{ted}, \cite{gp}, 
\cite{prd}, for collision inside the nonextremal horizon it is proximity of
the point of collision to the bifurcation surface \cite{inner}, for the
circular orbit in the background of the near-extremal black hole it is small 
$\kappa $, etc.

Now, we can add one more case to this number of cases. In the situation
discussed in the present work, where is no small parameter but another
factor prevents infinite $E_{c.m.}$. Either there is no critical trajectory
(the massive case) or, as an exception, general rule does not work and
collision does not give rise to infinite \thinspace $E_{c.m.}$ $\ $(a photon
moving along the horizon generator).

\section{Conclusion}

We have constructed a coordinate system regular on the horizon that
generalizes a similar system for the Kerr metric \cite{dor}.\ With its help,
it is shown that the circular near-horizon orbits of near-extremal black
holes asymptotically approach the generators of the horizon on the slices
where the new time $\bar{t}=const$ thus generalizing the observation made in 
\cite{ted11}.

In the extremal case, general approach to the description of circular
equatorial orbits near dirty black holes is suggested. It would seem that
there exist circular orbits with $r=r_{H}$ exactly on the horizon but such
orbits are fake for massive particles. Circular orbits are possible for
near-critical particles. Their radial distance to the horizon can be made as
small as one likes but, nonetheless, their radius does not coincide with
that of the horizon. If parameters of a black hole satisfy the exact
relation corresponding to a fake orbit on the horizon, this has physical
meaning despite the fact the orbit is fake. First, they correspond to the
value of the parameters (say, the angular momentum of \ a black hole $%
a=a_{0} $) such that around these values (when $a=a_{0}+a_{1}N_{0}+$...)
there exist circular orbits with small $N_{0}$. Second, these distance
values manifest itself in dynamics, when a fine-tuned (critical) particle
moves on orbits that are not exactly circular, slowly approaching the
horizon. When parameters of a black hole (say, its angular momentum) cross $%
a_{0}$, this results in the change of the rate with which a particle
approaches the horizon asymptotically.

The velocities measured by a local nonrotating observer are found on the
near-horizon orbits. In general, they do not coincide with the values
obtained in the extremal limits of near-extremal metrics. It is demonstrated
that properties of near-horizon circular orbits near extremal black holes
cannot be understood as the extremal limit of corresponding properties of
near-extremal black holes.

Connection with the BSW effect is revealed. The fact that on-horizon
circular orbits on the horizon are fake preserves the kinematic censorship
(the finiteness of energy in collisions) from violation.

It would be of interest to extend the approach and results of the present
work to the case of nonequatorial orbits.

\begin{acknowledgments}
This work was funded by the subsidy allocated to Kazan Federal University
for the state assignment in the sphere of scientific activities.
\end{acknowledgments}

\end{document}